\documentclass[a4paper,12pt]{article}
\begin{document}
\title{Rotation numbers of invariant manifolds around
 unstable periodic orbits for the diamagnetic Kepler problem}
\author{ Zuo-Bing Wu \\
State Key Laboratory of Nonlinear Mechanics, Institute of
Mechanics, \\
Chinese Academy of Sciences, Beijing 100080, China}
\maketitle

\begin{abstract}
In this paper, a method to construct topological template in terms of symbolic
dynamics for the diamagnetic Kepler problem is proposed.
To confirm the topological template, rotation numbers of
invariant manifolds around unstable periodic orbits in a phase space
are taken as an object of comparison.
The rotation numbers are determined from the definition
and connected with symbolic sequences
encoding the periodic orbits in a reduced Poincar\'e section.
Only symbolic codes with inverse ordering in the forward mapping
 can contribute to the rotation of invariant manifolds around the periodic orbits.
By using symbolic ordering, the reduced Poincar\'e section is constricted
along stable manifolds and a topological template,
which preserves the ordering of forward sequences
and can be used to extract the rotation numbers, is established.
The rotation numbers computed from the topological template are
the same as those computed from their original definition.
\end{abstract}

\newpage
\section{Introduction}

Many interesting nonlinear systems in experiments are wells
described by low-dimensional dynamical models. Their dynamical
processes include of transition of periodic orbits from stable to
unstable and bifurcation to chaos. To make a global understanding
of the system, some topological and geometrical methods based on
periodic orbits are developed[1]. Symbolic dynamics, as a
coarse-grained description of the dynamics, provides an effective
tool to depict the topological dynamics[2]. In special, for a
chaotic system, two curve families of stable and unstable
manifolds intersect each other and decompose a Poincar\'e
section[3,4]. Enumeration and existence of unstable periodic
orbits (UPOs)  in the Poincar\'e section are determined[5,6] and
some numerical methods such as finding UPOs are proposed[7].

Besides the features of dynamics in a Poincar\'e section,
evolution of manifolds in a phase space, which possibly gains a
geometric insight into the dynamics, is another important
characteristic of the system. For example, bifurcation to chaos
can be identified by tracing the evolution of stable and unstable
manifolds in the phase space with parameters. Recently, some
algorithms for computing two-dimensional stable and unstable
manifolds in a three-dimensional phase space are proposed[8,9,10]
and applied to visualize the structure of chaos[11]. Since UPOs
are closely related to invariant manifolds, a basic issue in their
relations is describing how local invariant manifolds rotate
around UPOs in the phase space. Moreover, how reducing their
topological relation into a two-dimensional template is important
for understanding the global organization of UPOs in chaos.

To motivate the visualization of rotation of invariant manifolds
around UPOs and construction of topological template in reduction
of stable manifolds, we consider the model for diamagnetic Kepler
problem (DKP)[12]. In our previous works, two coordinate axes are
chosen as a Poincar\'e section to form an annulus in a lifted
space. The dynamics on the annulus can be reduced by considering
the symmetry of system. In view of stretching and wrapping in the
lifted space, symbolic dynamics without involving bounces has been
established[13]. Due to the ordering of stable and unstable
manifolds in the minimal domain (a reduced Poincar\'e section), a
method to extract UPOs corresponding to short symbolic strings is
proposed. A one to one correspondence between UPOs and symbolic
sequences is shown under the system symmetry decomposition[14].
Although we only focused on the case of zero scaled energy, in our
numerical experiences, the methods can be still used at the scaled
energy $\pm 0.1$. In this paper, for the DKP, we calculate
rotation numbers of invariant manifolds around UPOs and set up a
connection between rotation numbers and symbolic sequences. Using
symbolic dynamics, we reduce stable manifolds to construct a
topological template, which preserves the topological relation of
UPOs.

\section{Model system and Poincar\'e section}
\label{sec:upo}

The Hamiltonian of a hydrogenic electron (with zero angular momentum)
in a uniform magnetic field ${\bf B}$
directed along the $z$-axis is given by
\begin{equation}
H=(1/2m) (p_{\rho}^2+p_z^2) -e^2/(\rho^2 +z^2)^{1/2} +\frac{1}{2} m\omega^2 \rho^2,
\label{1}
\end{equation}
where $\omega=eB/2mc$ is half the cyclotron frequency.
Converting to atomic units and
transforming the cylindrical coordinates to semiparabolic ones,
 the Hamiltonian becomes

\begin{equation}
h=\frac{p_\mu^2}{2}+\frac{p_\nu^2}{2}-\epsilon(\mu^2+\nu^2)+\frac{1}{8}\mu^2
\nu^2(\mu^2+\nu^2)
\equiv 2,
\label{3}
\end{equation}
where $\epsilon=E \gamma^{-2/3}$ is the scaled energy depending on
the energy $E$ and the dimensionless field strength parameter
$\gamma=2\omega$. The symmetry group of $h$ consists of the
identity $e$, two reflections $\sigma_{\mu}$, $\sigma_{\nu}$
across the $\mu$, $\nu$ axes, two diagonal reflections
$\sigma_{13}$, $\sigma_{24}$, and three rotations $C_4$, $C_2$ and
$C_4^3$ by $\pi/2$, $\pi$ and $3\pi/2$ around the center,
respectively[15]. In the following, the symmetries $C_4$, $C_2$
and $C_4^3$ are denoted by $\rho$, $\pi$ and $\bar{\rho}$,
respectively. The time-reversal symmetry is denoted by $T$.

Figure 1 displays an orbit and boundary of the transformed
potential for $\epsilon=0$. A Poincar\'e section is chosen as
follows. Imagine that the $\mu$ and $\nu$ axes are both of a
finite width and length. A counter-clockwise contour is taken
along the perimeter of the area forming by the two crossing
imaginary rectangles. The Poincar\'e section is then obtained by
recording the position and the tangent component of the momentum
along the contour, i.e., the Birkhoff canonical coordinates[16] at
intersecting points with the contour where an orbit enter the
inside of the contour. The length of the contour is infinite. It
is more convenient to transform the contour to one with a finite
length. For example, in the first quadrant, the transformations
$s=-\mu/(1+\mu)$ along the positive $\mu$ axis and $s=\nu/(1+\nu)$
along the positive $\nu$ axis convert the segment of the original
contour in the first quadrant to interval of length 2 parametrized
with $s \in [-1,1)$. The variable corresponding to the momentum is
taken as $v=-p_{\mu}/p$ at the positive $\mu$ axis, and
$v=p_{\nu}/p$ at the positive $\nu$ axis, where
$p=\sqrt{p_{\mu}^2+p_{\nu}^2}$. In this way we may parametrize the
whole contour with $s\in [-1,7)$ and define corresponding $v$. The
rotational symmetry under $\rho$, $\pi$ and $\bar{\rho}$ in the
original configurational space becomes the translational symmetry
of shifting $s$ by a multiple of 2 in the $s-v$ plane. The
dynamics on the Poincar\'e surface is then represented by a map on
the annulus $s\in[-1,7)$ and $v\in[-1,1]$, which is taken as a
fundamental domain (FD). When consider an image and preimage of
the FD, we need extend to its lifted space. The partial image and
preimage of the FD in the lifted space is given in Fig. 2. For
example, in Fig.~2(a), zones 1 (2) and 1' (2') in the strip 1 (2)
are mapped forward into zones +3 (+0) and +2 (+1), respectively.
In the same way, the backward mapping of the strip 1 is given in
Fig.~2(b). Since the rotational symmetry of the Hamiltonian (2)
corresponds to the
 translational symmetry in lifted space,
the annulus ($s\in[-1,7)$, $v\in[-1,1]$) on the Poincar\'e section
 can be reduced to a domain ($s\in[0,2)$, $v\in[-1,1]$).

In the conservative system, classical dynamics preserves an invariant volume in
the phase space under constricting, stretching and folding. This behavior can be displayed in the
Poincar\'e section. In Fig.~3(a), we draw the 9 lines ($s\in(0,1)$, $v\in[-1,1]$)
in the reduced domain (RD) and their forward mapping in the lifted space. In the mapping,
the original zone in the RD is stretched and folded, as well as wrapped.
In order to display ordering of the lines in the forward mapping, we also
plot connecting lines between two different strips in the lifted space.
In the lifted space ($s\in(2,3)$, $v\in[-1,1]$), the ordering of lines in the top-left
part preserves the same as the original one and the ordering in the bottom-right
part is in reverse.  In another lifted space ($s\in(1,2)$, $v\in[-1,1]$),
the ordering of lines in the whole part is in reverse.
It is clear that the wrapping of lines in the forward mapping is clockwise,
if the lines in the strip ($s\in(1,2)$, $v\in[-1,1]$) are stuck to those
in the strip ($s\in(2,3)$, $v\in[-1,1]$) in terms of their ordering.
In the same way, in Fig.~3(b), the similar result
can be obtained from the 9 lines ($s\in(1,2)$, $v\in[-1,1]$) in the RD and their forward mapping
in the lifted space.
So, the forward mapping illustrates the rotation of RD in the clockwise direction.

\section{Stable and unstable invariant manifolds}
\label{sec:manifolds}

In general, the invariant manifolds as a subset are contained in
manifolds. The method for calculating stable and unstable
manifolds (dynamical foliations) used in through the rest of our
works is detailed in [2]. Using the same method, we can thus
generate the stable and unstable invariant manifolds through
unstable periodic points in a two-dimensional Poincar\'e section.
Here we present a short introduction of the method.

(i) Unstable manifolds:  Taking a circle around $n$ steps backward mapping ($x_{-n}$, $y_{-n}$) of
an unstable periodic point ($x_0$, $y_0$), we get an ellipse centered at ($x_0$, $y_0$) after the
same steps forward mapping.
Its long axis points to the most stretching direction. When we fix the point ($x_0$, $y_0$) and increase
$n$, the ellipse is stretched and rotated, as well as its most stretching direction changes slightly.
When $n \rightarrow \infty$, the most stretching direction approaches a limit. This direction
is the most stretching direction of the point ($x_0$, $y_0$). After going a short distance along the direction,
we get a new point ($x_0$, $y_0$). Repeating the above process, we can get the new most
stretching direction. Finally, an unstable invariant manifold is generated by connecting the points ($x_0$, $y_0$).
In fact, the deformation of ellipse is closely related to the dynamical matrix of Poincar\'e
mapping.

(ii) Stable manifolds:  Taking a circle around $n$ steps forward mapping ($x_n$, $y_n$) of
an unstable periodic point ($x_0$, $y_0$), we get an ellipse centered at ($x_0$, $y_0$) after
the same steps backward mapping.
Following the above similar process, we can get the most stable direction of the point ($x_0$, $y_0$)
and a stable invariant manifold.

\section{Evolution of unstable manifolds in rotated Poincar\'e sections}

In order to display the evolution of invariant manifolds around an UPO,
we rotate counter-clockwise the ($\mu$, $\nu$) coordinates to the ($\mu_{\phi}$, $\nu_{\phi}$) coordinates
with an angle $\phi \in [0,\pi/2]$. By using the same transformations in Section II,
the Poincar\'e map ($s_{\phi}$, $v_{\phi}$) is obtained
from the ($\mu_{\phi}$, $\nu_{\phi}$) coordinates and reduced to a domain
($s_{\phi}\in[0,2)$, $v_{\phi}\in[-1,1]$).
Since $h$ has $C_{4v}$ and time-reversal symmetries, we take 4 UPOs (4)(5)(14)(15)
with different symmetries in the Table I
as examples to investigate the evolution of invariant manifolds.
Their plots in the configuration
space with 10 rotation coordinate axes and periodic points with unstable invariant manifolds
in the RDs are drown in Figs.~4(a)-(d), respectively.
From the initial point in the $+\nu_{\phi=0}$ coordinate axis corresponding to
the point 1 in the RD with $\phi=0$, each UPO goes into the second quadrant as displayed
in the configuration space by an arrow.  Its time process is recorded
in the RDs with 10 rotation angles $\phi$.  So, the figure in the RD with $\phi=\pi/2$
is the same as that in the RD with $\phi=0$ besides the first point in the former RD
is the second point in the later one. In a periodic process, we will calculate
the advanced phase $\theta$ of an unstable manifold in rotation around the periodic orbits
to determine rotation number $\theta/2\pi$.
The phase $\theta$ is counted positive (negative) when the rotation of unstable manifold
around the periodic orbits is counter-clockwise (clockwise).

The UPOs with 4 different symmetries are described as follows:

(i) The UPO (4) displayed in the configuration space of Fig.~4(a)
has $\sigma_{\mu}$ and $T$ symmetries, but not $\rho$, $\pi$, $\bar{\rho}$,
$\sigma_{\nu}$, $\sigma_{13}$ and $\sigma_{24}$ symmetries.
In the second quadrant of configuration space, the orbit starting from
the $+\nu_{\phi=0}$ coordinate axis
goes to the $-\mu_{\phi=0}$ ($+\nu_{\phi=\pi/2}$) coordinate axis.
It corresponds to that the point 1 moves to 2
in the RD with $\phi=0$, i.e. 1 in the RD with $\phi=\pi/2$.
In the process, the orbit passes through
the $+\nu_{\phi=\pi/18}$, $+\nu_{\phi=\pi/9}$, $\cdots$, $+\nu_{\phi=4\pi/9}$
coordinate axes, while the point 1 moves in the RDs with $\phi=\pi/18$,
$\phi=\pi/9$, $\cdots$, $\phi=4\pi/9$. At the same time, an unstable manifold
passing through the point 1 evolves in the RDs. The phase of unstable manifold
advances by an angle close to $-\pi$.
In the third quadrant of configuration space, the orbit starting
from the $-\mu_{\phi=0}$ coordinate axis goes to the $-\nu_{\phi=0}$ ($-\mu_{\phi=\pi/2}$)
coordinate axis. It corresponds to that the point 2  moves to 3 in the RD
with $\phi=0$, i.e. 2 in the RD with $\phi=\pi/2$.
In the process, the orbit passes through
the $-\mu_{\phi=\pi/18}$, $-\mu_{\phi=\pi/9}$, $\cdots$, $-\mu_{\phi=4\pi/9}$
coordinate axes, while the point 2 moves in the RDs with $\phi=\pi/18$,
$\phi=\pi/9$, $\cdots$, $\phi=4\pi/9$. The phase of unstable
manifold advances by an angle close to $-\pi$.
The orbit starting from the $-\nu_{\phi=0}$ coordinate axis
goes into the fourth quadrant of configuration space and back to the $-\nu_{\phi=0}$
coordinate axis. It corresponds to that the point 3 moves to 4
in the RD with $\phi=0$. In the process, the orbit passes through the
$-\nu_{\phi=\pi/18}$ coordinate axis two times,
while the point 3 goes into the RD with $\phi=\pi/18$,
then leaps to the point 4 and comes back the RD with $\phi=0$. In the moving processes
of the point 3 from the RD with $\phi=0$ into the RD with $\phi=\pi/18$ and
of the point 4 from the RD with $\phi=\pi/18$ into the RD with $\phi=0$, the directions of
unstable manifolds are almost invariant. However, in the leaping process of the orbit from the point 3
to 4 in the RD with $\phi=\pi/18$, an unstable direction indicated as the arrow
from the periodic point to its neighboring point on the unstable manifold
is approximately reversed. The phase of unstable manifold advances
by an angle close to $-\pi$.
In the third quadrant of configuration space, the orbit starting
from the $-\nu_{\phi=0}$ ($-\mu_{\phi=\pi/2}$) coordinate axis goes to the $-\mu_{\phi=0}$
coordinate axis. It corresponds to that the point 4 moves to 5 in the RD
with $\phi=0$, i.e., the point 3 in the RD with $\phi=\pi/2$ moves to 5 in the RD
with $\phi=0$.  In the process, the orbit passes through
the $-\mu_{\phi=4\pi/9}$, $-\mu_{\phi=7\pi/18}$, $\cdots$, $-\mu_{\phi=\pi/18}$
coordinate axes, while the point 3 moves in the RDs with $\phi=4\pi/9$,
$\phi=7\pi/18$, $\cdots$, $\phi=\pi/9$ and then the point 5 moves in the RD with $\phi=\pi/18$.
Since the $-\nu_{\phi=\pi/18}$ coordinate axis intersects the orbit in the fourth quadrant,
two points 3 and 4 are added in the RD with $\phi=\pi/18$.
The phase of unstable manifold advances by an angle close to $-\pi$.
In the second quadrant of configuration space, the orbit starting from
the $-\mu_{\phi=0}$ ($+\nu_{\phi=\pi/2}$) coordinate axis
goes to the $+\nu_{\phi=0}$ coordinate axis. It corresponds to that
the point 5 moves to 6 in the RD with $\phi=0$, i.e., the point 4
in the RD with $\phi=\pi/2$ moves to 6 in the RD with $\phi=0$.
In the process, the orbit passes through
the $+\nu_{\phi=4\pi/9}$, $+\nu_{\phi=7\pi/18}$, $\cdots$, $+\nu_{\phi=\pi/18}$
coordinate axes, while the point 4 moves in the RDs with $\phi=4\pi/9$,
$\phi=7\pi/18$, $\cdots$, $\phi=\pi/9$ and then the point 6 moves in the RD with $\phi=\pi/18$.
The phase of unstable manifold advances by an angle close to $-\pi$.
The orbit starting from the $+\nu_{\phi=0}$ ($+\mu_{\phi=\pi/2}$) coordinate axis goes into
the first quadrant of configuration space and back to the $+\nu_{\phi=0}$ ($+\mu_{\phi=\pi/2}$)
coordinate axis. It corresponds to that the point 6 moves to 1
in the RD with $\phi=0$, i.e., the point 5 moves to 6 in the RD with $\phi=\pi/2$.
 In the process, the orbit passes through the $+\mu_{\phi=4\pi/9}$ coordinate axis two times,
 while the point 5 goes into the RD with $\phi=4\pi/9$, then
leaps to 6 and comes back the RD with $\phi=\pi/2$. In the moving processes
of the point 5 from the RD with $\phi=\pi/2$ into the RD with $\phi=4\pi/9$ and
of the point 6 from the RD with $\phi=4\pi/9$ into the RD with $\phi=\pi/2$, the directions of
unstable manifolds are almost invariant. However, in the leaping process from the point 5
to 6 in the RD with $\phi=4\pi/9$, an unstable direction
is approximately reversed. The phase of unstable manifold advances
by an angle close to $-\pi$.

   So, in the periodic process, the unstable manifold returns
to its original position, as well as the phase of unstable manifold advances by -6$\pi$.
The rotation number of UPO (4) is -3.

(ii) The UPO (5) displayed in the configuration space of Fig.~4(b)
passes through the origin. Its right limit orbit
has $\sigma_{13}$ symmetry, but not $\rho$, $\pi$, $\bar{\rho}$, $T$,
$\sigma_{\mu}$, $\sigma_{\nu}$ and $\sigma_{24}$ symmetries.
Similarly, in the periodic process,
the phase of unstable manifold advances by -6$\pi$.
The rotation number of UPO (5) is -3.

(iii) The UPO (14) displayed in the configuration space of Fig.~4(c)
has $\sigma_{\nu}$ symmetry, but not $\rho$, $\pi$, $\bar{\rho}$, $T$,
$\sigma_{\mu}$, $\sigma_{13}$ and $\sigma_{24}$ symmetries.
Similarly, in the periodic process,
the phase of unstable manifold advances by -8$\pi$.
The rotation number of UPO (14) is -4.

(iv) The UPO (15) displayed in the configuration space of Fig.~4(d)
has $\rho$, $\pi$, $\bar{\rho}$ symmetries, but not $T$, $\sigma_{\mu}$,
$\sigma_{\nu}$, $\sigma_{13}$ and $\sigma_{24}$ symmetries.
Similarly, in the periodic process, the phase of unstable manifold advances by -16$\pi$.
The rotation number of UPO (15) is -8.

    Thus, in the rotation of unstable invariant manifolds around UPOs,
we have determined rotation numbers. At the same time, we have also
obtained that the advanced phase $|\theta|$ of unstable invariant manifold in a Poincar\'e mapping
does not exceed $\pi$. In the Sect. V, we will present a method to
calculate the rotation numbers in a Poincar\'e section.

\section{Rotation of unstable manifolds in a Poincar\'e section}

According to the natural ordering in the lifted space and the
occurrence of tangencies of manifolds, we have the region
partition in the RD with symbols ($L_0$, $R_0$, $R_1$, $R_2$ and
$L_2$) and the ordering for forward sequences[13]

\begin{equation}
\bullet L_0< \bullet R_0< \bullet R_1< \bullet R_2< \bullet L_2.
\label{md}
\end{equation}
The forward mapping preserves the ordering in regions of $\bullet L_0$ and $\bullet L_2$, but
reverses the ordering in regions of $\bullet R_0$, $\bullet R_1$ and $\bullet R_2$.

In the RD, some symmetries of the Hamiltonian (2) are reduced, it can be reflected
by the relation of orbit periods to sequence ones. So,
we can firstly calculate the advanced phases of unstable directions in rotation around UPOs
 in the RD and then add the contribution of symmetries to determine rotation numbers.
 The 4 UPOs with different symmetries in the Sect. IV are still taken as
examples.

(i) In Fig.~5(a), we draw the periodic points encoded by $R_0^2R_1R_2^2R_1$ and
stable and unstable invariant manifolds passing through the points.
The periodic points are denoted by circles. In order to illustrate
the evolution of unstable direction around the periodic points, we take another initial point
near the periodic point 1 on the unstable invariant manifold.
The forward mapping of the point is also drawn in the figure and denoted by crosses.
The arrows from periodic points to their neighboring points on the unstable invariant manifolds
display unstable directions.
In the forward mapping from the periodic point 1 to 2, the symbolic sequence $\bullet R_0^2R_1R_2^2R_1$
is shifted to $\bullet R_0R_1R_2^2R_1R_0$ and the original unstable direction is approximately
reversed. Since the rotation is clockwise, we obtain $-\pi$ rotation
of the unstable direction. In the forward mapping from the periodic point 2 to 3,
the symbolic sequence $\bullet R_0R_1R_2^2R_1R_0$
is shifted to $\bullet R_1R_2^2R_1R_0^2$ and the unstable direction is approximately
reversed. We also obtain $-\pi$ rotation of the unstable direction,
i.e. $-2 \pi$ rotation of the original unstable direction. In the same way,
$-\pi$ rotation of the unstable direction is obtained
in the forward mapping from the periodic point 3 to 4.
In the periodic point 4, we take another neighboring point denoted by a triangle
to replace the point denoted by a cross. In the same way,
$3 \times (-\pi)$ rotation of the unstable direction is obtained in the forward mapping
from the periodic point 4 to 5, from the periodic point 5 to 6
and from the periodic point 6 to 1.
Thus, during the mapping in the sequence period, the original unstable direction goes back
and the total advance of phase is $-6 \pi$. Since the orbit period is equal to
the sequence one, i.e., the UPO (4) has not the $\rho$, $\pi$, $\bar{\rho}$ symmetries,
 the rotation number of UPO encoded by $R_0^2R_1R_2^2R_1$ is -3.

(ii) In Fig.~5(b),
the periodic points with the right limit encoded by $L_0R_0^2R_1L_2R_2^2R_1$
and the  stable and unstable invariant manifolds passing through the points are drawn.
In the forward mapping from the  periodic point 1 to 2, the symbolic sequence $\bullet L_0R_0^2R_1L_2R_2^2R_1$
is shifted to $\bullet R_0^2R_1L_2R_2^2R_1L_0$ and the original unstable direction is approximately
preserved. In the forward mapping from the  periodic point 5 to 6, the same result is obtained.
In other forward mappings, $6 \times (-\pi)$ rotation of the unstable direction is added.
Thus, during the mapping in the sequence period, the original unstable direction goes back
and the total advance of phase is $-6 \pi$. Since the orbit period is equal to
the sequence one, the rotation number of UPO (5)
encoded by $L_0R_0^2R_1L_2R_2^2R_1$ is -3.

(iii)In Fig.~5(c), the periodic points encoded by $L_0R_1R_2R_1^2L_2R_1R_0R_1^2$
 and stable and unstable invariant manifolds passing through the points are drawn.
In the forward mapping from the  periodic point 1 to 2, the symbolic sequence
$\bullet L_0R_1R_2R_1^2L_2R_1R_0R_1^2$ is shifted to $\bullet R_1R_2R_1^2L_2R_1R_0R_1^2L_0$
and the original unstable direction is approximately preserved.
In the forward mapping from the  periodic point 6 to 7, the same result is obtained.
In other forward mappings, $8 \times (-\pi)$ rotation of the unstable direction is added.
Thus, during the mapping in the sequence period, the original unstable direction goes back
and the total advance of phase is $-8 \pi$. Since the orbit period is equal to
the sequence one, the rotation number of UPO (14)
encoded by $L_0R_1R_2R_1^2L_2R_1R_0R_1^2$ is -4.

(iv) In Fig.~5(d), the periodic points encoded by $L_0R_1^2R_0^2$
 and stable and unstable invariant manifolds passing through the points are drawn.
In the forward mapping from the  periodic point 1 to 2, the symbolic sequence
$\bullet L_0R_1^2R_0^2$ is shifted to $\bullet R_1^2R_0^2L_0$
and the original unstable direction is approximately preserved.
In other forward mappings, $4 \times (-\pi)$ rotation of the unstable direction is added.
Thus, during the mapping in the sequence period, the original unstable direction goes back
and the total advance of phase is $-4 \pi$. Since the orbit period is 4 times of
the sequence one, i.e., the UPO (15) has the $\rho$, $\pi$, $\bar{\rho}$ symmetries,
the rotation number of the UPO
encoded by $L_0R_1^2R_0^2$ is -8.

In the above examples describing the rotation of unstable directions around periodic points,
the forward map corresponding to the shift with $L_0$ or $L_2$ ($R_0$ or $R_1$ or $R_2$) approximately
preserves (reserves) the original unstable direction.
So, we can multiply the
numbers of $R_0$, $R_1$ and $R_2$ in 5-letter symbolic sequences by one half of the ratios of
orbit periods to sequence ones to determine rotation numbers of UPOs.

Since the RD has the $\pi$-rotation symmetry, the 5-letter
symbolic dynamics can be reduced to the 3-letter one in the
minimal domain (MD) ($s \in [0,1)$, $v \in [-1,1]$)[13]. The MD is
partitioned and denoted by symbols $L_0$, $R_0$ and $R_1$. The
correspondence of 5-letter symbolic sequences with 3-letter ones
is $L_0 \rightarrow L_0$, $R_0 \rightarrow R_0$, $R_1 \rightarrow
R_1$, $R_2 \rightarrow R_0$ and $L_2 \rightarrow L_0$. In general,
the number of $R_0$, $R_1$ and $R_2$ in 5-letter symbolic
sequences is twice of the number of $R_0$ and $R_1$ in 3-letter
ones. The ratios of orbit periods to sequence ones for the former
are one half of those for the later. Of course, the simple
repeating of 3-letter symbolic sequences in 5-letter ones will be
removed. For example, the 3-letter symbolic sequences $R_0^2R_1$
and $L_0R_0^2R_1$ correspond to the 5-letter ones
$R_0(R_2)R_0(R_2)R_1R_2(R_0)R_2(R_0)R_1$ and
$L_0(L_2)R_0(R_2)R_0(R_2)R_1L_2(L_0)R_2(R_0)R_2(R_0)R_1$,
respectively. Rotation numbers of two UPOs can be determined by
calculating total numbers of letters $R_0$ and $R_1$ in 3-letter
symbolic sequences and multiplying them by one half of the ratios
of orbit periods to sequence ones. The same rotation numbers of
the UPOs can be also obtained by using the method for 5-letter
symbolic sequences. Thus, using the method, we extract rotation
numbers of 38 UPOs from symbolic sequences as given in Table I.

\section{Topological Template}

After the region partition and symbolic ordering are introduced,
the families of stable and unstable manifolds constitute curve
coordinates in the RD. Each stable (unstable) manifold has the
same forward (backward) symbolic sequence. The ordering on stable
(unstable) manifolds is described by that of forward (backward)
symbolic sequences[14]. In Fig.~6, two families of sub-manifolds
divided by the partition line $\bullet C_0$ or $\bullet C_2$ have
the opposite ordering. The ordering of stable (unstable) manifolds
increases monotonically from the left-bottom (left-top) to
right-top (right-bottom) along each unstable (stable) manifold.

Along each stable manifold in zones $\bullet L_0$, $\bullet R_0$ and $\bullet R_1$
of Fig.~6,
we constrict all points in the curve to a point. The point preserves the forward
symbolic sequence and the ordering of stable manifold. So, in the left region of Fig.~6,
the points in three zones are reduced to three lines. Connecting the three lines,
we obtain a belt partitioned by the symbols $\bullet C_0$ and $\bullet B_0$,
and denoted by the symbols $\bullet L_0$, $\bullet R_0$ and $\bullet R_1$ as given
in the top of Fig.~7(a). From the left to right along the belt, the ordering of forward
sequences increases monotonically.
The forward mapping of the left region ($s \in (0,1)$ and $v \in [-1,1]$) in Fig.~6,
i.e. the right region ($s \in (2,3)$, $v \in [-1,1]$) in Fig.~3(a), can be reduced in the RD.
So, in the forward mapping, the zones $\bullet L_0$ and $\bullet R_0$
still keep in the region ($s \in (0,1)$, $v \in [-1,1]$) encoded by $L_0 \bullet$ and $R_0 \bullet$,
respectively, but the zone $\bullet R_1$ moves in the region ($s \in (1,2)$, $v \in [-1,1]$) encoded
by $R_1 \bullet$.
In Fig.~6, we again partition the RD and encode it by corresponding backward symbols.
The two zones $L_0 \bullet$ and $R_0 \bullet$ are partitioned to five zones
by the lines $\bullet B_0$ and $\bullet C_0$, as well as the zone $R_1 \bullet$
is partitioned to three zones by the lines $\bullet B_2$ and $\bullet C_2$.
In the stretching and folding processes of forward mapping,
the original three zones are mapped into the eight zones.
We still constrict all points along stable manifolds in each zone.
Thus, the points in eight zones are reduced to five lines in the left region
and three lines in the right region.
Connecting the eight lines, we obtain a belt partitioned
by the symbols $L_0 \bullet B_0$, $C_0 \bullet R_1$, $R_0 \bullet B_0$, $R_0 \bullet C_0$,
$R_1 \bullet C_2$ and $R_1 \bullet B_2$,
and encoded by the strings $L_0 \bullet R_0$, $L_0 \bullet R_1$, $R_0 \bullet R_1$,
$R_0 \bullet R_0$, $R_0 \bullet L_0$, $R_1 \bullet L_2$,
$R_1 \bullet R_2$ and $R_1 \bullet R_1$ in the bottom of Fig.~7(a).
The ordering of the bottom belt is the same as that of the top one.
So, the forward mapping of the region ($s \in (0,1), v \in [-1,1]$)
can be described by a twisting part of topological template as given in Fig.~7(a).

In the same way, along each stable manifold in zones $\bullet R_1$, $\bullet R_2$
and $\bullet L_2$ of Fig.~6, all points in the curve  are constricted to a point.
A belt containing the point is partitioned by the symbols $\bullet B_2$ and $\bullet L_2$,
and encoded by the symbols $\bullet R_1$, $\bullet R_2$ and $\bullet L_2$ as given in the top of Fig.~7(b).
From the left to right along the belt, the ordering of forward sequences increases monotonically.
The forward mapping of the right region ($s \in (1,2)$ and $v \in [-1,1]$) in Fig.~6,
i.e. the left region ($s \in (-1,0)$, $v \in [-1,1]$) in Fig.~3(b), can be reduced in the RD.
So, in the forward mapping,
the zones $\bullet L_2$ and $\bullet R_2$ still keep in the region ($s \in (1,2)$, $v \in [-1,1]$)
encoded by $L_2 \bullet$ and $R_2 \bullet$,
but the zone $\bullet R_1$ moves in the region ($s \in (0,1)$, $v \in [-1,1]$) encoded by $R_1 \bullet$.
We still constrict all points along stable manifolds in each zone
and obtain the belt partitioned
by the symbols $R_1 \bullet B_0$, $R_1 \bullet C_0$, $R_2 \bullet C_2$, $R_2 \bullet B_2$,
$C_2 \bullet R_1$ and $L_2 \bullet B_2$,
and encoded by the strings $R_1 \bullet R_1$, $R_1 \bullet R_0$, $R_1 \bullet L_0$,
$R_2 \bullet L_2$, $R_2 \bullet R_2$, $R_2 \bullet R_1$, $L_2 \bullet R_1$ and
$L_2 \bullet R_2$ in the bottom of Fig.~7(b).
 So, the forward mapping
of the region ($s \in (1,2)$, $v \in [-1,1]$) can be described by a twisting part of topological
template as given in Fig.~7(b).

In the two twisting parts of topological template, the belts reflect approximately
the direction of unstable manifolds.
Using the twisting parts of topological templates,
we can easily calculate rotation numbers for given symbolic sequences. In the same way,
the 4 UPOs with different symmetries in Sect. II are still taken as examples.
For the sequence $R_0^2R_1R_2^2R_1$ encoding the UPO(4), firstly, $R_0 \bullet R_0R_1R_2^2R_1$
is obtained by shifting $R_1 \bullet R_0^2 R_1 R_2^2$. When an arrow is put on the $\bullet R_0$
zone of top belt in Fig.~7(a), after a forward mapping,
the arrow  is moved on the $R_0 \bullet R_0$ zone of bottom belt in Fig.~7(a).
Since the arrow rotates clockwise to its opposite direction, we count
the process as -1. Then, $R_0 \bullet R_1R_2^2R_1R_0$ is obtained by shifting
$R_0\bullet R_0R_1R_2^2R_1$. Since an arrow on the $\bullet R_0$ zone of
 top belt in Fig.~7(a)
rotates clockwise to its opposite direction, we count the
forward mapping as -1. Repeating the above process, we get the
total number -6 counting the forward mapping in sequence period.
Since the orbit period is equal to the sequence one, we can thus obtain
rotation number of the UPO encoded by $R_0^2R_1R_2^2R_1$ is -3.
For the sequence $L_0R_0^2R_1L_2R_2^2R_1$ encoding the UPO(5),
after a forward mapping for $\bullet L_0R_0^2R_1L_2R_2^2R_1$,
an arrow on the top belt in Fig.~7(a) moves parallelly
on the zone $L_0\bullet R_0$ of bottom belt.
We count the forward mapping as 0. Following the same process, we
get the total number -6 counting the forward mapping in the sequence period.
Since the orbit period is equal to the sequence one, we can thus obtain
rotation number of the UPO encoded by $L_0R_0^2R_1L_2R_2^2R_1$ is -3.
Similarly, for the sequence $L_0R_1R_2R_1^2L_2R_1R_0R_1^2$ ($L_0R_1^2R_0^2$)
encoding the UPO(14) (UPO(15)),
we get the total number -8 (-4) counting the forward mapping in the sequence period.
Since the orbit period is equal to (4 times of) the sequence one.
 we can thus obtain rotation number of the UPO encoded by $L_0R_1R_2R_1^2L_2R_1R_0R_1^2$ ($L_0R_1^2R_0^2$)
is -4 (-8).
By comparing with the former computation for the 4 UPOs from the definition and in a Poincar\'e section,
the same results are extracted from the topological template.

By combining the two twisting parts in Figs.~7(a)(b), suspension of the Poincar\'e mapping,
which displays the relative position of zones in the forward mapping from the top belt to bottom
one, is obtained. A global topological template of the RD is constructed by connecting the
suspension with a flow corresponding to the Poincar\'e mapping in Fig.~8.
The template preserves the ordering of forward sequences encoding stable manifolds in a belt
and the same or inverse ordering of symbolic encoding in the forward mapping
of all parts in the belt.

\section{Conclusion and discussion}
\label{sec:sum}

In summary, we have presented the systematic study of
the evolution of invariant manifolds around unstable periodic orbits and
the reduction of them to construct a topological template
in terms of symbolic dynamics for the diamagnetic Kepler problem.
To confirm the topological template,
rotation numbers of invariant manifolds around unstable periodic
orbits in a phase space, which quantify the evolution, are determined from the definition
 and connected with symbolic sequences encoding the periodic orbits.
Only symbolic codes, which correspond
to the forward mapping with inverse ordering,
 can contribute to the rotation of invariant manifolds.
By using symbolic ordering, the reduced Poincar\'e section is constricted along stable manifolds
and a topological template, which preserves the ordering of forward sequences
and can be used to extract the rotation numbers, is established.
The rotation numbers computed from the topological template are
the same as those computed from the original definition.

Since unstable periodic orbits in phase space are the skeleton of the chaotic system,
the local evolution of manifolds near unstable periodic orbits
can present basic features of the global evolution of manifolds in phase space.
One of the basic features can be quantified by the rotation number and embedded in
the topological template.

In the semiclassical Green's function, the phase correction is
related to Maslov indices of the UPOs[17], which has been
connected with symbolic sequences of unstable periodic orbits due
to boundary coding[18]. The relation of Maslov indices to rotation
numbers of unstable periodic orbits remains to be determined.

\newpage
\section{REFERENCES}

[1] J. Guckenheimer and P. Holmes,  Nonlinear Oscillations,
Dynamical Systems and Bifurcation of Vector Fields, Springer, New
York, 1983.

[2] B.-L. Hao and W.-M. Zheng, Applied Symbolic Dynamics and
Chaos, World Scientific, Singapore, 1998.

[3] P. Grassberger and H. Kantz, Phys. Lett. A {\bf 113} (1985)
235.

[4] Y. Gu, Phys. Lett. A {\bf 124}, (1987) 340.

[5] P. Cvitanovi\'c, G. H. Gunaratne and I. Procaccia, Phys. Rev.
A {\bf 38}, (1988) 1503.

[6] P. Grassberger, H. Kantz and U. Moenig, J. Phys. A {\bf 22},
(1989) 5217.

[7] K. T. Hansen, Phys. Rev. E {\bf 52}, (1995) 2388.

[8] M. E. Johnson, M. S. Jolly, and I. G. Kevrekidis, Numer.
Algorithms {\bf 74}, (1997) 125.

[9] M. Dellnitz and A. Hohmann, Numer. Algorithms {\bf 75}, (1997)
293.

[10] B. Krauskopf and H. Osinga, Chaos {\bf 9}, (1999) 768.

[11] H. M. Osinga and B. Krauskopf, Computers and Graphics {\bf
26}, (2002) 815.

[12] H. Friedrich and D. Wintgen, Phys. Repts. {\bf 183}, (1989)
37.

[13] Z.-B. Wu and W.-M. Zheng, Physica Scripta {\bf 59}, (1999)
266 [$chao-dyn/9907016$].

[14] Z.-B. Wu and J.-Y. Zeng, Physica Scripta {\bf 61}, (2000) 406
[$nlin/0004004$].

[15] P. Cvitanovi\'c and B. Eckhardt, Nonlinearity {\bf 6}, (1993)
277.

[16] G. D. Birkhoff, Acta Mathematica {\bf 50}, (1927) 359.

[17] J. B. Delos, Advan. in Chem. Phys. {\bf 65}, 161 (1986).

[18] B. Eckhardt and D. Wintgen, J. Phys. B {\bf 23}, 355 (1990).

\newpage
\begin{table}
\begin{footnotesize}
Table I. Rotation numbers and symbolic sequences of UPOs for the
diamagnetic Kepler problem at $\epsilon=0$.
\begin{tabular}{crcrccc}
No& 3-lett. Seq. \& & its Period  & 5-lett. Seq. \& & its Period & Orb. Period& Rot. Num.\\
1   & $R_0$ & 1     & $R_0$&1&4     &   -2\\
2   & $L_0R_1$ & 2      & $L_0R_1L_2R_1$&4&4    &   -1\\
3   & $R_0R_1$ & 2      & $R_0R_1R_2R_1$&4&4    &   -2\\
4   & $R_0^2R_1$ & 3        & $R_0^2R_1R_2^2R_1$&6&6    &   -3\\
5   & $L_0R_0^2R_1$ & 4 & $L_0R_0^2R_1L_2R_2^2R_1$&8&8  &   -3\\
6   & $L_0R_1R_0^2$ & 4 & $L_0R_1R_2^2L_2R_1R_0^2$&8&8  &   -3\\
7   & $L_0R_1R_0R_1$ & 4    & $L_0R_1R_2R_1$&4&4    &   -1.5\\
8   & $R_0^3R_1$ & 4        & $R_0^3R_1R_2^3R_1$&8&8    &   -4\\
9   & $R_0^2R_1^2$ & 4      & $R_0^2R_1^2$&4&8  &   -4\\
10  & $L_0R_0L_0R_1^2$& 5   & $L_0R_0L_0R_1^2$&5&20 &   -6\\
11  & $L_0R_0^2R_1^2$& 5    & $L_0R_0^2R_1^2$&5&20  &   -8\\
12  & $L_0R_1L_0R_1^2$& 5   & $L_0R_1L_2R_1^2L_2R_1L_0R_1^2$&10&10  &   -3\\
13  & $L_0R_1R_0^2R_1$& 5   & $L_0R_1R_2^2R_1$&5&20 &   -8\\
14  & $L_0R_1R_0R_1^2$& 5   & $L_0R_1R_2R_1^2L_2R_1R_0R_1^2$&10&10  &   -4\\
15  & $L_0R_1^2R_0^2$& 5    & $L_0R_1^2R_0^2$&5&20  &   -8\\
16  & $L_0R_1^2R_0R_1$& 5   & $L_0R_1^2R_0R_1L_2R_1^2R_2R_1$&10&10  &   -4\\
17  & $R_0^4R_1$& 5     & $R_0^4R_1R_2^4R_1$&10&10  &   -5\\
18  & $R_0^3R_1^2$  & 5 & $R_0^3R_1^2$&5&20 &   -10\\
19  & $R_0^2R_1R_0R_1$& 5   & $R_0^2R_1R_2R_1$&5&20 &   -10\\
20  & $R_0^2R_1^3$  & 5 & $R_0^2R_1^3R_2^2R_1^3$&10&10  &   -5\\
21  & $R_0R_1R_0R_1^2$& 5   & $R_0R_1R_2R_1^2R_2R_1R_0R_1^2$&10&10  &   -5\\
22  & $L_0R_0L_0R_1R_0R_1$& 6   & $L_0R_0L_0R_1R_2R_1$&6&12 &   -4\\
23  & $L_0R_0L_0R_1^3$& 6   & $L_0R_0L_0R_1^3L_2R_2L_2R_1^3$&12&12  &   -4\\
24  & $L_0R_0^2L_0R_1^2$& 6 & $L_0R_0^2L_0R_1^2$&6&6    &   -2\\
25  & $L_0R_0^4R_1$ & 6 & $L_0R_0^4R_1L_2R_2^4R_1$&12&12    &   -5\\
26  & $L_0R_0^3R_1^2$& 6    & $L_0R_0^3R_1^2$&6&6   &   -2.5\\
27  & $L_0R_0^2R_1^3$& 6    & $L_0R_0^2R_1^3L_2R_2^2R_1^3$&12&12    &   -5\\
28  & $L_0R_1L_0R_1R_0R_1$& 6   & $L_0R_1L_2R_1R_0R_1L_2R_1L_0R_1R_2R_1$&12&12  &   -4\\
29  & $L_0R_1R_0^4$ & 6 & $L_0R_1R_2^4L_2R_1R_0^4$&12&12    &   -5\\
30  & $L_0R_1R_0^3R_1$& 6   & $L_0R_1R_2^3R_1$&6&12 &   -5\\
31  & $L_0R_1R_0R_1R_0R_1$& 6   & $L_0R_1R_2R_1R_0R_1L_2R_1R_0R_1R_2R_1$&12&12  &   -5\\
32  & $L_0R_1^2R_0^3$& 6    & $L_0R_1^2R_0^3$&6&6   &   -2.5\\
33  & $L_0R_1^3R_0^2$& 6    & $L_0R_1^3R_2^2L_2R_1^3R_0^2$&12&12    &   -5\\
34  & $R_0^5R_1$& 6     & $R_0^5R_1R_2^5R_1$&12&12  &   -6\\
35  & $R_0^4R_1^2$  & 6 & $R_0^4R_1^2$&6&12 &   -6\\
36  & $R_0^3R_1R_0R_1$& 6   & $R_0^3R_1R_2R_1$&6&12 &   -6\\
37  & $R_0^3R_1^3$& 6       & $R_0^3R_1^3R_2^3R_1^3$&12&12  &   -6\\
38  & $R_0^2R_1^4$  & 6 & $R_0^2R_1^4$&6&12 &   -6\\
\end{tabular}
\end{footnotesize}
\end{table}

\newpage
\section{FIGURE CAPTION}

Fig.~1. A typical orbit and boundary of the transformed potential
for the diamagnetic Kepler problem at $\epsilon=0$.

Fig.~2. A image (a) and preimage (b) of the strips 1 and 2 of the
fundamental domain ($s\in[-1,7)$, $v\in[-1,1]$) in the lifted
space.

Fig.~3.  9 lines in (a) ($s\in(0,1)$, $v\in[-1,1]$) or (b)
 ($s\in(1,2)$, $v\in[-1,1]$) of the reduced domain
and their forward mapping in the correspondent lifted space. The
different types of lines display the relative changes between
original positions and their forward mappings along the $s$
coordinate axis.

Fig.~4. UPOs with different symmetry in the configuration space
and periodic points with unstable invariant manifolds in rotation
Poincar\'e sections: (a) the UPO (4); (b) the UPO (5); (c) the UPO
(14); (d) the UPO (15).

Fig.~5. Periodic points encoded by (a) $R_0^2R_1R_2^2R_1$; (b)
$L_0R_0^2R_1L_2R_2^2R_1$; (c) $L_0R_1R_2R_1^2L_2R_1R_0R_1^2$; (d)
$L_0R_1^2R_0^2$ in the 5-letter encoding and their stable and
unstable invariant manifolds.

Fig.~6. Stable and unstable manifolds with partition lines in the
reduced domain.

Fig.~7. Twisting parts of topological template describing the
forward mapping on the regions (a) ($s \in [0,1)$, $v \in [-1,1]$)
and (b) ($s \in [1,2)$, $v \in [-1,1]$). For crossing of two lines
in suspension of the Poincar\'e mapping, the front (back) one is
denoted by solid lines (short dashes or the combination of solid
lines and short dashes). The top and bottom belts are denoted by
solid lines or dashes depending on their positions. The connecting
lines of two parts of the broken belt in the forward mapping are
denoted by long dashes.

Fig.~8. A flow of topological template of the reduced domain. The
notions for belts are the same as Fig.~7, except that projection
of several parts in the bottom belt on one belt in terms of
forward symbolic codings is connected by short dashes. The flow is
denoted by short dashes.

\end{document}